# Shear-induced droplet mobility within porous surfaces

Si Suo[1, 3], Haibo Zhao[1], Shervin Bagheri[3], Peng Yu[1 a *] and Yixiang Gan[2,4 b *]

[1] Department of Mechanics and Aerospace Engineering, Southern University of Science and Technology, China

[2] School of Civil Engineering, The University of Sydney, Australia

[3] Department of Mechanics, KTH Royal Institute of Technology, 100 44 Stockholm, Sweden.

[4] The University of Sydney Nano Institute (Sydney Nano), The University of Sydney, NSW 2006, Australia

[a] yup6@sustech.edu.cn; [b] yixiang.gan@sydney.edu.au.

**Abstract**

Droplet mobility under shear flows is important in a wide range of engineering applications, e.g., fog collection, and self-cleaning surfaces. For structured surfaces to achieve superhydrophobicity, the removal of stains adhered within the microscale surface features strongly determines the functional performance and durability. In this study, we numerically investigate the shear-induced mobility of the droplet trapped within porous surfaces. Through simulations covering a wide range of flow conditions and porous geometries, three droplet mobility modes are identified, i.e., the stick-slip, crossover, and slugging modes. To quantitatively characterise the droplet dynamics, we propose a droplet-scale capillary number that considers the driving force and capillary resistance. By comparing against the simulation results, the proposed dimensionless number presents a strong correlation with the leftover volume. The dominating mechanisms revealed in this study provide a basis for further research on enhancing surface cleaning and optimising design of anti-fouling surfaces.

**Graphical abstract:**

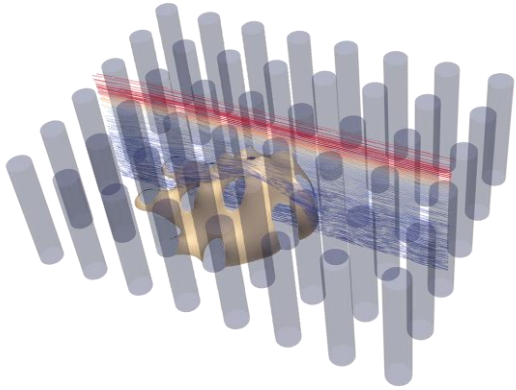 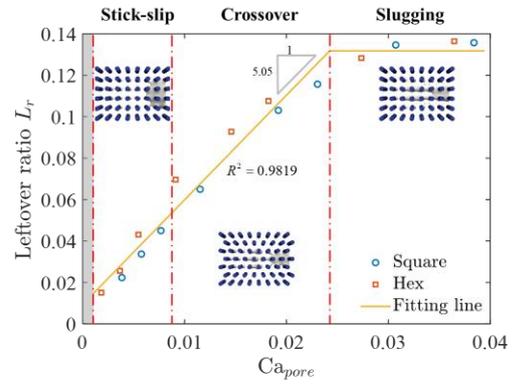

# 1. Introduction

Understanding droplet deformation, spreading, pinning, splitting etc., is fundamental to engineering processes, e.g., water management in fuel cells [1, 2], enhanced oil recovery [3], and surface cleaning and coating [4, 5]. The basic setting of shear flow passing a droplet adhering a solid wall has been extensively investigated through experiments and numerical simulations. Before the onset of motion on a smooth wall, the resting droplet deforms under the shear flow, until it reaches a critical stage with increasing shear rate. Numerical studies indicated that the steady-state droplet shape is a consequence of the competition among viscous, capillary and inertial effects [6, 7]. The critical conditions can be characterised by Weber number, i.e., the ratio between inertial and capillary force, as a function of Reynolds number and contact angle hysteresis [8, 9]. Similar conclusions were also reported from the theoretical analysis [10] and experimental observations [4, 11]. Once a droplet is initiated to move, a variety of motion patterns are observed depending on the various combinations of fluid and surface properties. It was experimentally revealed that with increasing droplet viscosity and surface wettability the droplet tends to be *slugging*, i.e., from moving as a whole to sliding with a trailing tail behind the droplet [12]. As a consequence of larger droplet volume, it is possible to observe that the droplet may be lifted off the surface following the deformation and sliding [11, 13], i.e., droplet pinch-off and detachment. External factors, e.g., surfactants [14] or electricity fields [15], may also influence the droplet behaviours.

Water management plays an essential role in performance and long-term maintenance of fuel cells [16, 17]. During the fuel cell operation, the water produced from electrochemical reactions is transported and accumulated on the gas diffusion layers (GDLs) forming distinct droplets. The liquid removal mainly depends on the droplet movement driven by the air stream. This flow process has been studied widely [18-20], especially regarding the impact of surface roughness on the wetting and detachment of the droplets adhered on GDLs. Specifically, the roughness can alter the apparent surface wettability, i.e., the rougher the surface is, the more likely that the droplet touches GDLs in the Cassie mode rather than the Wenzel mode and therefore promoting detachment [21]; additionally, the roughness may affect the moving trail of a droplet resulting in deviation from the channel centre and touching the channel walls [22]. On the other hand the microstructure of GDLs can be artificially designed to assist the droplet motion [23].

Understanding how the surface roughness and porosity impact droplet mobility not only contributes to optimising design of fuel cells but is also relevant for enhanced oil recovery at a much larger scale. Traditionally, oil recovery depends on the flooding displacement from fractured reservoir [24-26]. However, interfacial instabilities, such as, viscous or capillary fingering, can result in undesired oil entrapping within pore space [27, 28]. Further collecting trapped oil phase is key to improve the recovery efficiency. Thus, gaining further knowledge of ganglia motion within flow paths containing complex surfaces is necessary.

In the recent decades, a variety of artificial textured surfaces have been proposed for anti-icing [29], self-cleaning [30], superhydrophobicity [31, 32], etc. Current research on droplet mobility mainly focuses on two aspects; droplet impact on solid surfaces and its post-impact effects including splashing, spreading, bouncing, sticking, gas trapping, etc [33, 34]; and spontaneous or programmable liquid movement along designed pathways [35, 36], even anti-gravity climbing [37, 38]. To push the artificial surfaces forward in practice, we still need comprehensive knowledge regarding the droplet-surface wetting processes. Especially, when the functional surfaces are exposed to complex environment, such as varying temperature and humidity, liquid may be trapped within surficial microstructures, as shown in Figure 1, which could impede the surface functionality. Therefore, understanding the dynamics of trapped liquid is especially important to liquid-repellant surface and the long-term maintenance of functional surfaces. Regarding droplet-porous matrix interactions, the focus of current research efforts are within the regimes of high mobility, e.g., bubbles flowing through saturated granular beds [39, 40], high speed, e.g., droplet impacting on or through porous media [41-43], or quasi-static state, e.g., droplet settling on the porous surfaces [44-46]. However, to the authors' knowledge, there is few studies directly related to the mobility of droplet trapped within porous surfaces.

To address this need, we investigate the process of a droplet trapped in the porous surface being displaced under shear flows through 3D pore-scale simulations. We focus on droplets whose characteristic length is smaller than the surface feature size, and thus *within* the surface structure. In the Darcy-flow regime, the droplet behaviour is determined by the competition between viscous and capillary effects. We demonstrate a transition of droplet motion mode, i.e., from capillary-dominated stick-slip mode to viscous-dominated slugging mode. A pore-scale capillary number is proposed by considering the tortuosity of obstacle arrangements to characterise the mode transition. By a correlation analysis on the leftover ratio, we discover that the drain-out left ratio is approximately proportional to the pore-scale capillary number and reaches a maximum value when slugging occurs. The current work sheds light on the droplet - porous media interaction in the Darcy-flow regime, and the discovered mechanism of droplet motion transition may pave the way to the enhanced surface cleaning.

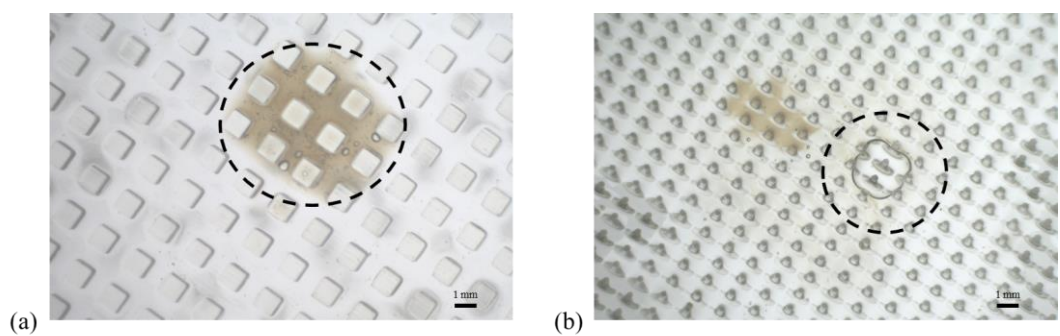

Figure 1. Trapped liquid stain within the artificial surfaces, marked by black dash circles on patterned surface structures with: (a) square pillars, and (b) round pillars.

## 2. Numerical model

*Numerical method*

In this study, we adopt the volume of fluid (VOF) to capture the droplet dynamics as it has been proved efficient and effective in pore-scale simulation [47-49]. For incompressible two-phase flows, the governing equations, including the continuity equation (1), phase fraction equation (2), and momentum equation (3), are shown as

$$\nabla \cdot \boldsymbol{u} = 0, \tag{1}$$

$$\frac{\partial \alpha}{\partial t} + \nabla \cdot (\boldsymbol{u}\alpha) = 0, \tag{2}$$

$$\frac{\partial (\rho \boldsymbol{u})}{\partial t} + \nabla \cdot (\rho \boldsymbol{u} \cdot \boldsymbol{u}) - \nabla \cdot (\eta \nabla \boldsymbol{u}) - (\nabla \boldsymbol{u}) \cdot \nabla \eta = -\nabla p + \mathbf{f}_\sigma. \tag{3}$$

Here $\alpha$ is the phase fraction of two fluids, $\boldsymbol{u}$ represents the velocity field, $\rho$ and $\eta$ represent the weighted average of density and viscosity, respectively, i.e., $\rho = \alpha \rho_{f1} + (1-\alpha)\rho_{f2}$ and $\eta = \alpha \eta_{f1} + (1-\alpha)\eta_{f2}$, and $p$ is the pressure. The surface tension $\mathbf{f}_\sigma$ is modelled as a continuum surface force [50],

$$\mathbf{f}_\sigma = \sigma \kappa \nabla \alpha, \tag{4}$$

where $\sigma$ is the surface tension and $\kappa$ is the local interface curvature related to the normal direction of the interface $\boldsymbol{n}_\alpha$, i.e.,

$$\boldsymbol{n}_\alpha = \frac{\nabla \alpha}{|\nabla \alpha|}, \text{ and} \tag{5}$$

$$\kappa = \nabla \cdot \boldsymbol{n}_\alpha. \tag{6}$$

The partial wetting conditions on solid surface can be applied by adjusting the interface normal,

$$\boldsymbol{n}_\alpha = \boldsymbol{n}_s \cos(\theta_w) + \boldsymbol{\tau}_s \sin(\theta_w), \tag{7}$$

where $\boldsymbol{n}_s$ and $\boldsymbol{\tau}_s$ are the normal and tangential vectors of solid surfaces, respectively, and $\theta_w$ is the equilibrium contact angle. Regarding the implementation of partially wetting conditions and inlet-outlet boundaries, one can refer to [51]. The numerical solution of Eqs. (1~3) plus boundary conditions is obtained by using OpenFoam, an open source CFD toolbox.

As shown in Figure 2, the simulations are conducted in a 3D domain with patterned obstacles to represent the porous surface, which are arranged in two forms, i.e., square and hexagon arrays. By adjusting obstacle radius, both arrangements possess the same porosity and throat size. The geometric parameters of two arrangements are summarised in Table 1. A pressure difference $\Delta P$ is applied between the left (inlet) and right (outlet) boundaries, and the top face of the computation domain is set to be an open boundary. Regarding velocity boundaries, zero-flux conditions are applied. Surfaces of the bottom and obstacles are wettable no-slip walls with a given constant contact angle $\theta_w$ of 60°. The droplet is initially placed close to the inlet with radius $r_d$ of 3 mm.

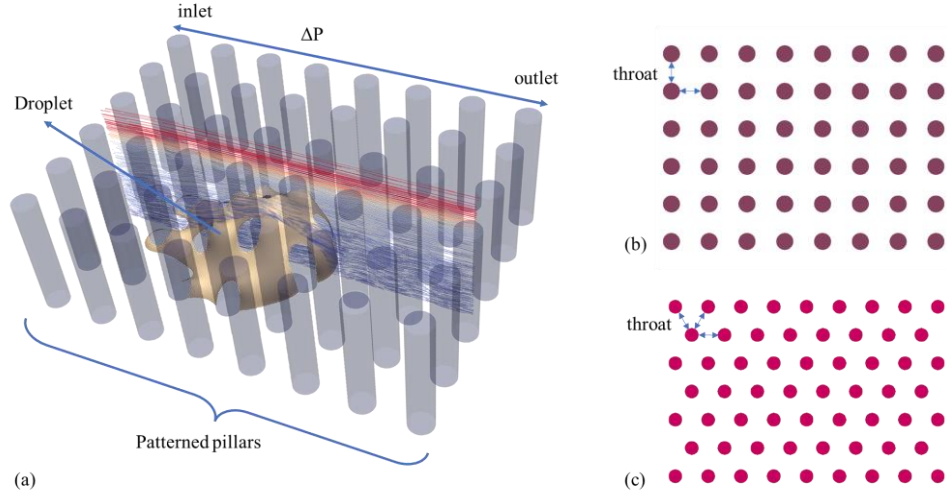

Figure 2. (a) The schematic of pore-scale numerical model, i.e., a trapped droplet is driven by pressure difference $\Delta P$ from inlet to outlet and moves through a porous surface, whose solid space is composed of patterned round pillars arranged in two forms, i.e., (b) square array and (c) hexagon array.

Table 1. Geometric parameters of the square and hex arrangement

|  | porosity $\phi$ [/] | throat size $r_t$ [mm] | obstacle radius $r_r$ [mm] | tortuosity $\tau$ [/] | permeability $K$ [mm$^2$] |
|---|---|---|---|---|---|
| Square | 0.84 | 1.4 | 0.56 | 1.0179 | 39.9 |
| Hex | 0.84 | 1.4 | 0.50 | 1.1715 | 16.5 |

The pioneering works regarding the droplet mobility [52, 53] suggested that the behaviour of droplet is determined by the combined effects involving capillary, viscous and inertial forces. Thus, to comprehensively investigate the mobility of a trapped droplet within porous surfaces, besides the varied porous geometries as above, the driving pressure and surface tension cover a wide range, i.e., $\Delta P = 100 \sim 2000$ Pa and $\sigma = 36 \sim 144$ mN/m. As for fluid properties, the ambient and droplet fluids are assumed to have the same viscosity $\eta_A = \eta_D = 1$ mPa·s, and density $\rho_A = \rho_D = 1 \times 10^3$ kg/m³.

*Mesh sensitivity*

To balance the numerical accuracy and computational cost, we design the meshing scheme for the computation domain as shown in Figure 3(a). Specifically, the whole computational domain is divided into three zones, i.e., the coarse zone on the top covering the free-surface flow domain, where the mesh size $\Delta x_c$ is regarded as a reference and fixed as 0.17 mm. The refined zone on the bottom of the porous domain covering the droplet moving region, where the mesh size $\Delta x_r$ is refined compared with $\Delta x_c$. These two zones are bridged by a transition zone. An index, leftover ratio $L_r$, defined as a ratio of the total leftover liquid volume inside the computation domain to the initial droplet volume, is used to measure the convergence of mesh-dependence study. Three mesh schemes, i.e., mesh 1 ($\Delta x_r = \Delta x_c/2$), mesh 2 ($\Delta x_r = \Delta x_c/4$), mesh 3 ($\Delta x_r = \Delta x_c/8$) are tested and compared, as shown in Figure 3(b). The results of mesh 2 and 3 overlap, and therefore mesh 2 is adopted in the following simulations.

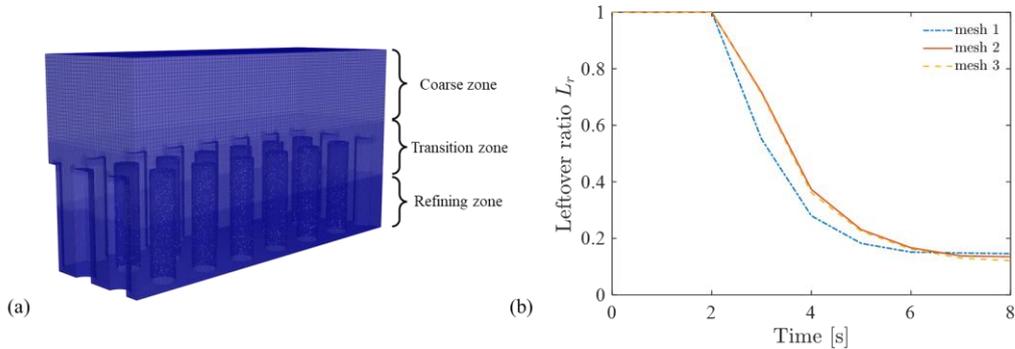

Figure 3. (a) The meshing scheme of the numerical model, including a coarse zone, a refined zone and a transition zone. (b) The comparison on the leftover ratio among three meshing schemes under the flow condition of $\Delta P = 1000$ Pa and $\sigma = 72$ mN/m.

## 3. Results and discussion

For a shear-flow driven droplet adhered to a solid surface, three modes of motion, namely sliding, crawling and detachment, can be observed depending on surface properties and flow conditions [11, 12]. However, the mobility of a moving droplet trapped in a porous surface is more complex since the flow dynamics and interface morphology are strongly modified by the solid structures. The process involves droplet splitting, attaching, and spreading, as a result of the interaction between viscous and capillary effects. Within a limited domain (17.5 mm × 12.5 mm), the process can be generally divided into pre- and post-drain-out stages. The moment when the rear of a droplet leaves the outlet is defined as the "drain-out" state. In this study, we focus on the droplet motion during the pre-drain-out stage.

To characterize the shear-induced droplet motion at the pre-drain-out stage, the capillary number, defined as the ratio of viscous force to capillary force, is adopted [54]. The capillary force can be represented by the surface tension force $\sigma r_d$, while the viscous force can be estimated from the characteristic flow velocity, i.e., $\eta_A \overline{U} r_d$. In porous media, Darcy's law is generally adopted to estimate the flow rate $\overline{U}$, as a field-average characterization for a large domain, assuming the flow is slow enough that inertial effect can be neglected. In this work, the Reynolds number $\text{Re} = \frac{\rho_A U r_r}{\eta_A} \leq 0.65$ in all simulations and therefore the above assumption is satisfied. According to Darcy's law, the field-average flow velocity $U_{\text{Darcy}}$ is

$$U_{\text{Darcy}} = \frac{K}{\eta_A} \frac{\Delta P}{L}, \tag{8}$$

where $L$ is the length of the domain from the inlet to the outlet. However, compared with the field-scale length, the droplet size is much smaller and comparable to the pore or throat size. Therefore, the droplet motion is controlled by the local flow field and a pore-scale quantity can better work as a characteristic flow rate. Here, the tortuosity is introduced to further modify the $U_{\text{Darcy}}$ as a pore-scale average flow rate $U_{\text{pore}}$ [55],

$$U_{\text{pore}} = \frac{U_{\text{Darcy}}}{\phi} \tau. \tag{9}$$

We adopt $U_{\text{pore}}$ as the characteristic velocity $\bar{U}$, and finally the pore-scale capillary number is obtained as

$$\text{Ca}_{\text{pore}} = \frac{\eta_A U_{\text{pore}}}{\sigma}, \tag{10}$$

which is adopted to describe the interaction between capillary and viscous effects during droplet motion for the following analysis.

## 3.1 Simulation results

By using the numerical model presented in Section 2, simulations of the trapped droplet moving through a porous surface are performed under various flow and geometric conditions. The results suggest that the moving processes can be divided into three motion modes, i.e., the stick-slip, crossover, and slugging mode, as shown in Figure 4. For the *stick-slip* mode, the droplet behaviour is controlled by the capillary effects, i.e., $\text{Ca}_{\text{pore}}$ is smaller than 0.01. The droplet almost retains its initial shape and the interface is displaced pore-by-pore. Noticeably, as marked by yellow circles in Figure 4(a), when the rear interface crosses the obstacles, a suction force induced by the curved interface is exerted on the meniscus. The direction, which is indicated by a red arrow marked in Figure 4(a), leads to droplet splitting and partial liquid attaching to the obstacles as a leftover component. Unlike the stick-slip mode, the *slugging* mode is dominated by the viscous effects, i.e., $\text{Ca}_{\text{pore}}$ is comparably larger than 0.02. Specifically, the droplet spreads as a thin film and its interface moves much faster along the main channels than in the down-wind pore space due to the increased viscous resistance. As a consequence, liquid tails are left behind the droplet and wrapped around obstacles, as marked by red circles in Figure 4(c), which may further shed smaller droplet attached obstacles. Bridging the stick-slip and slugging modes, the *crossover* mode is observed around the balance point of viscous and capillary effects, resulting in the mixed features of the other two modes. The kinematics of these motion modes are further illustrated in the supplementary animations.

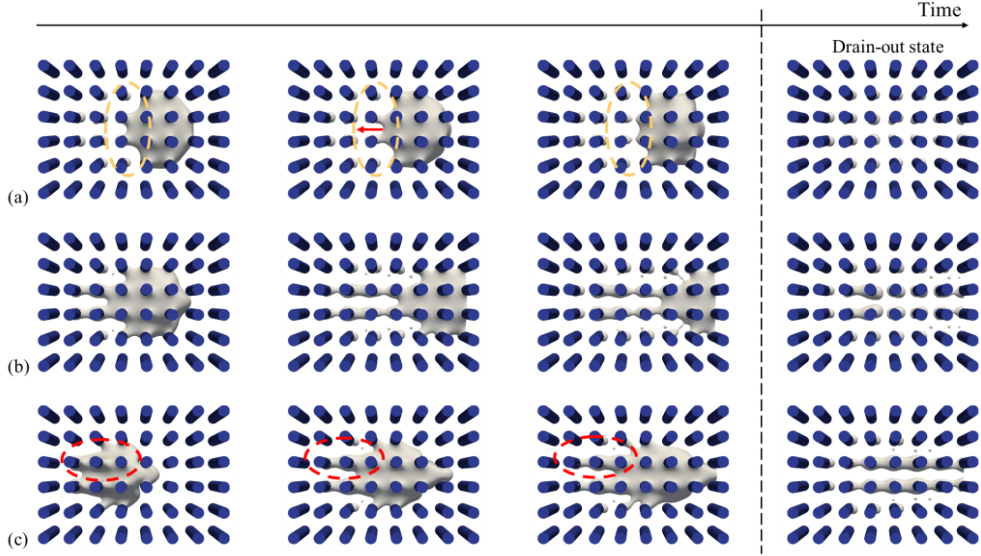

Figure 4. The evolution of three droplet motion modes in square arrangement: (a) stick-slip mode ($\Delta P = 200$ Pa, $\sigma = 72$ mN/m, $\text{Ca}_{\text{pore}} = 0.0077$), (b) crossover mode ($\Delta P = 500$ Pa, $\sigma = 72$ mN/m, $\text{Ca}_{\text{pore}} = 0.0192$), and (c) slugging mode ($\Delta P = 1000$ Pa, $\sigma = 72$ mN/m, $\text{Ca}_{\text{pore}} = 0.0384$). The last column indicates the liquid distribution of each mode at the drain-out state.

Once the droplet leaves the region of interest from the outlet, part of it remains within the porous surfaces in the form of smaller drops or long slugs sticking to the obstacles. The distribution of the remaining liquid volume of drops or slugs can further characterise the motion modes. Figure 5 shows the cumulative distribution function (CDF) $F$ of remaining drop volumes, defined as a proportion of drops smaller than a given volume $\bar{V}_d$ normalized by the initial droplet volume. For the stick-slip mode, the droplet marches slowly and stably like a quasi-static process so that the droplet is split evenly and the liquid volume distributes within a narrow range, i.e., the remaining drops share the similar size. By contrast, the liquid volumes of the slugging mode cover a much wider range because the droplet spreads fast and long slugs occur. In crossover mode, the CDF mostly lies in the middle since the small drops and short slugs co-exist.

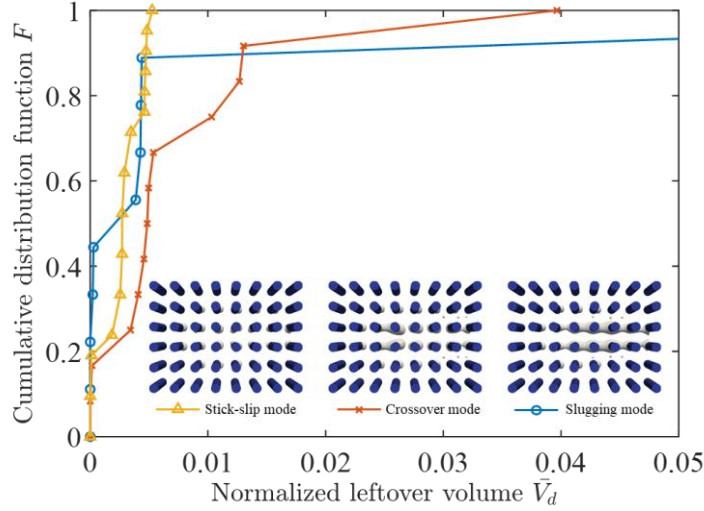

Figure 5. Cumulative distribution function of dispersed liquid leftover volumes at the drain-out state, which are normalised by the initial droplet volume, for three motion modes.

We can use the index $L_r$ to describe the displacing process. The change of $L_r$ is directly related to the switch of droplet motion modes, i.e., it is expected to be small for the stick-slip mode and large for the slugging mode. Simulation results, as demonstrated in Figure 4, suggest that conditions that are related to the viscous and capillary effects can trigger the switch of droplet motion modes and finally contribute to the leftover ratio. Specifically, with the driving pressure increasing, as shown in Figure 6(a), the viscous effect is enhanced, i.e., $\mathrm{Ca_{pore}}$ increases from 0.0038 to 0.0384, and thus droplet motion experiences a transition from the stick-slip to slugging mode as indicated in Figure 4. Correspondingly, the drain-out leftover ratio increases more than 10%. Also, the average droplet moving speed, which can be estimated by the drain-out time, increases with driving pressure. The surface tension directly contributes to the capillary effects, and with its increment as shown in Figure 6(b) from I ($\mathrm{Ca_{pore}} = 0.0230$), II ($\mathrm{Ca_{pore}} = 0.0115$) to III ($\mathrm{Ca_{pore}} = 0.0058$), the droplet motion tends to follow the stick-slip mode. In general, both the viscous and capillary effects depend on the geometry. The capillary force is mainly determined by the porosity and throat size [46], which are the same for the two configuration (see Table 1). The viscous effects on the other hand are different for the two different arrangements, due to the significant difference on the permeability of both configurations (see Table 1). The effective flow rate of the square arrangement is larger than that of the hex arrangement under the same driving pressure and correspondingly larger for $\text{Ca}_{\text{pore}}$ in the square arrangement so that the droplet motion within the square arrangement is closer to the slugging mode, as shown in Figure 6(c).

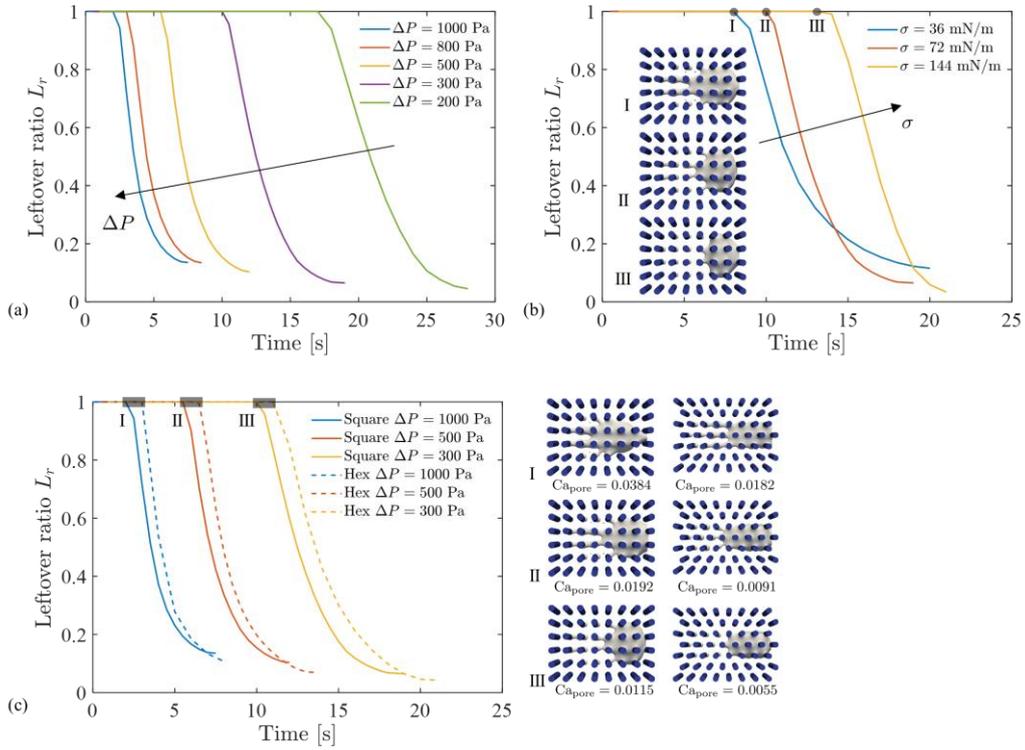

Figure 6. The evolution of leftover ratio $L_r$ from the onset to the drain-out state for different cases: (a) square arrangement with $\sigma = 72$ mN/m and $\Delta P = 200$, 300, 500, 800 and 1000 Pa; (b) square arrangement, $\Delta P = 300$ Pa, $\sigma = 36$, 72 and 144 mN/m; (c) comparison between square and hex arrangement with $\sigma = 72$ mN/m and $\Delta P = 300$, 500 and 1000 Pa.

In summary, the competition between capillary and viscous effects controls the behavior of a trapped droplet moving through a porous surface, and correspondingly determines the liquid leftover which could be a substantial index in patterned surface optimization and flow process design, especially for surface cleaning engineering. So, we conducted a correlation analysis as follows to build a quantitative description of drain-out state leftover ratio.

## 3.2 Correlation analysis

The drain-out state leftover ratio $L_r$ as a function of $\text{Ca}_{\text{pore}}$ is shown in Figure 7. At the first stage, covering the stick-slip and crossover mode, $L_r$ monotonically increases with $\text{Ca}_{\text{pore}}$ and approximately follows linear relationship, i.e.,

$$L_r = \lambda \text{Ca}_{\text{pore}} + L_r^0, \tag{11}$$

where $\lambda$ is the scale coefficient and $L_r^0$ is the vertical intercept. The simulation data for both configurations can collapse on this fitting line as shown in Figure 7, and the function of $\text{Ca}_{\text{pore}}$ to $L_r$ matches the proposed model in Eq. (11) with $R^2 = 0.9819$. Then, when $\text{Ca}_{\text{pore}}$ is beyond 0.0230, $L_r$ enters a plateau which indicates that the slugging mode occurs and the maximum volume of droplet are trapped as long slugs. If the driving pressure further increases, inertial effects may be involved, and the droplet pinch-off is expected, which is beyond the scope of this study. Noticeably, if the viscous force is too small to overcome the capillary barrier, corresponding to a threshold value of $\text{Ca}_{\text{pore}} \leq 0.0015$, marked by the grey area in Figure 7, the droplet is stationary due to pinning.

It should be noted that the proposed pore-scale capillary number, $\text{Ca}_{\text{pore}}$, unifies the results from different array types considered in this study across a wide range of physical parameter space. The index can be used to guide and optimise the surface cleaning processes in practice for given fluid properties and surface structures.

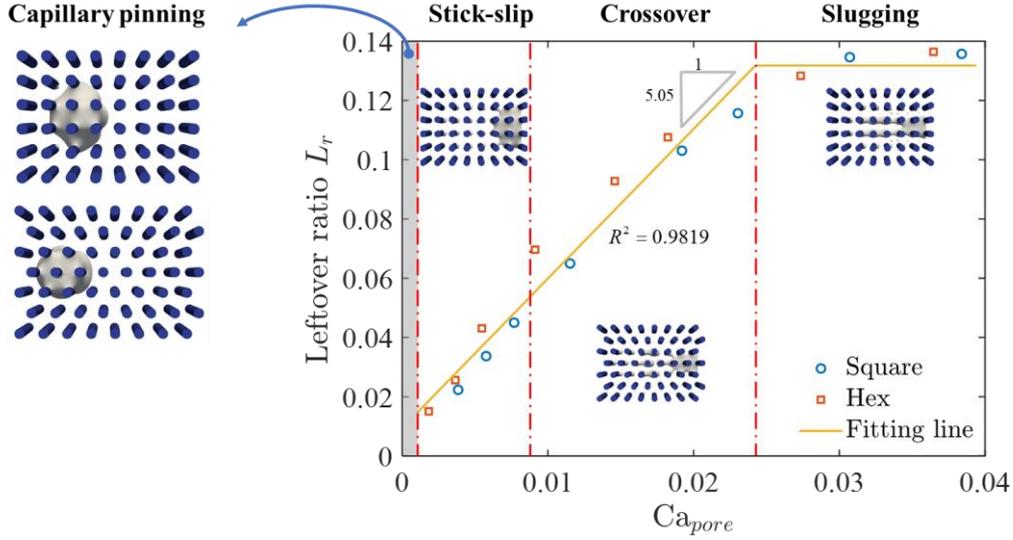

Figure 7. Drain-out state leftover ratio $L_r$ vs. pore-scale capillary number $\mathrm{Ca}_{\mathrm{pore}}$. The mode boundaries are marked by the red dot-dash line, as $\mathrm{Ca}_{\mathrm{pore}} \approx 0.0015$, $0.0090$ and $0.0230$.

## 4. Conclusion

In this work, we focus on the dynamics of a trapped droplet moving through porous surfaces, and the combined impacts of viscous and capillary forces on droplet mobility are numerically investigated by pore-scale simulations covering various flow conditions and geometric configurations. Specifically, three droplet motion modes are identified. With the continuous enhancement in the viscous effects relative to the capillary effects, the moving pattern transits from the stick-slip, crossover to slugging mode and the remaining liquid forms separate drops, short slugs to long slugs, correspondingly. The droplet mobility and transition of its motion modes are a consequence of the competition between viscous and capillary effects, and therefore can be naturally characterised by the proposed capillary number $\mathrm{Ca}_{\mathrm{pore}}$, which captures the pore scale flow conditions. Furthermore, we find a strong correlation between $\mathrm{Ca}_{\mathrm{pore}}$ and the drain-out state leftover ratio $L_r$, which presents two stages, i.e., $L_r$ linearly increases with $\text{Ca}_{\text{pore}}$ at the first stage covering the stick-slip and crossover modes; and $L_r$ reaches the maximum value and enters a platform once the long slugs occurs at $\text{Ca}_{\text{pore}} \approx 0.0230$. The conclusion of this study can provide guidance on optimising design of patterned surfaces, especially for the anti-fouling features as well as surface cleaning improvement. Furthermore, knowledge on the inertial-involved processes is worth exploring as an extension of this work since it may benefit the engineering involving liquid atomisation and droplet dispersion, etc.

**Acknowledgement**

This work was supported by the National Natural Science Foundation of China (Grant No. 12172163 and 91852205).

# References

1. Mukherjee, P.P., C.-Y. Wang, and Q. Kang, *Mesoscopic modeling of two-phase behavior and flooding phenomena in polymer electrolyte fuel cells.* Electrochimica Acta, 2009. **54**(27): p. 6861-6875.
2. Bazylak, A., D. Sinton, and N. Djilali, *Dynamic water transport and droplet emergence in PEMFC gas diffusion layers.* Journal of Power Sources, 2008. **176**(1): p. 240-246.
3. Wang, F., et al., *Dynamic analysis of deformation and start-up process of residual-oil droplet on wall under shear flow.* Journal of Petroleum Science and Engineering, 2021. **199**: p. 108335.
4. Thoreau, V., et al., *Physico-chemical and dynamic study of oil-drop removal from bare and coated stainless-steel surfaces.* Journal of adhesion science and technology, 2006. **20**(16): p. 1819-1831.
5. Christodoulou, C., et al., *Mathematical modelling of water absorption and evaporation in a pharmaceutical tablet during film coating.* Chemical Engineering Science, 2018. **175**: p. 40-55.
6. Dimitrakopoulos, P. and J.J.L. Higdon, *On the displacement of three-dimensional fluid droplets from solid surfaces in low-Reynolds-number shear flows.* Journal of Fluid Mechanics, 1998. **377**: p. 189-222.
7. Dimitrakopoulos, P., *Deformation of a droplet adhering to a solid surface in shear flow: onset of interfacial sliding.* Journal of Fluid Mechanics, 2007. **580**: p. 451.
8. Spelt, P.D., *Shear flow past two-dimensional droplets pinned or moving on an adhering channel wall at moderate Reynolds numbers: a numerical study.* Journal of Fluid Mechanics, 2006. **561**(439): p. 8.
9. Ding, H. and P.D.M. Spelt, *Onset of motion of a three-dimensional droplet on a wall in shear flow at moderate Reynolds numbers.* Journal of Fluid Mechanics, 2008. **599**: p. 341-362.
10. Dussan V., E.B., *On the ability of drops to stick to surfaces of solids. Part 3. The influences of the motion of the surrounding fluid on dislodging drops.* Journal of Fluid Mechanics, 1987. **174**(-1): p. 381-397.
11. Seevaratnam, G.K., et al., *Laminar flow deformation of a droplet adhering to a wall in a channel.* Chemical Engineering Science, 2010. **65**(16): p. 4523-4534.
12. Fan, J., M.C.T. Wilson, and N. Kapur, *Displacement of liquid droplets on a surface by a shearing air flow.* Journal of Colloid and Interface Science, 2011. **356**(1): p. 286-292.
13. Madani, S. and A. Amirfazli, *Oil drop shedding from solid substrates by a shearing liquid.* Colloids and Surfaces A: Physicochemical and Engineering Aspects, 2014. **441**: p. 796-806.
14. Liu, H., et al., *Modelling a surfactant-covered droplet on a solid surface in three-dimensional shear flow.* Journal of Fluid Mechanics, 2020. **897**.
15. Raman, K.A., et al., *Electrically induced droplet ejection dynamics under shear flow.* Physics of Fluids, 2020. **32**(3): p. 032103.
16. Theodorakakos, A., et al., *Dynamics of water droplets detached from porous surfaces of relevance to PEM fuel cells.* Journal of Colloid and Interface Science, 2006. **300**(2): p. 673-687.

17. Kumbur, E.C., K.V. Sharp, and M.M. Mench, *Liquid droplet behavior and instability in a polymer electrolyte fuel cell flow channel.* Journal of Power Sources, 2006. **161**(1): p. 333-345.
18. Niblett, D., et al., *Two-phase flow dynamics in a gas diffusion layer - gas channel - microporous layer system.* Journal of Power Sources, 2020. **471**: p. 228427.
19. Jeon, D.H. and H. Kim, *Effect of compression on water transport in gas diffusion layer of polymer electrolyte membrane fuel cell using lattice Boltzmann method.* Journal of Power Sources, 2015. **294**: p. 393-405.
20. Xu, A., W. Shyy, and T. Zhao, *Lattice Boltzmann modeling of transport phenomena in fuel cells and flow batteries.* Acta Mechanica Sinica, 2017. **33**(3): p. 555-574.
21. Bao, Y. and Y. Gan, *Roughness effects of gas diffusion layers on droplet dynamics in PEMFC flow channels.* International Journal of Hydrogen Energy, 2020. **45**(35): p. 17869-17881.
22. Hou, Y., et al., *3D lattice Boltzmann modeling of droplet motion in PEM fuel cell channel with realistic GDL microstructure and fluid properties.* International Journal of Hydrogen Energy, 2020. **45**(22): p. 12476-12488.
23. Chen, L., Y.-L. He, and W.-Q. Tao, *Effects of surface microstructures of gas diffusion layer on water droplet dynamic behaviors in a micro gas channel of proton exchange membrane fuel cells.* International Journal of Heat and Mass Transfer, 2013. **60**: p. 252-262.
24. Akai, T., et al., *Modeling Oil Recovery in Mixed-Wet Rocks: Pore-Scale Comparison Between Experiment and Simulation.* 2018.
25. Nilsson, M.A., et al., *Effect of fluid rheology on enhanced oil recovery in a microfluidic sandstone device.* Journal of Non-Newtonian Fluid Mechanics, 2013. **202**: p. 112-119.
26. Yun, W., et al., *Toward Reservoir-on-a-Chip: Rapid Performance Evaluation of Enhanced Oil Recovery Surfactants for Carbonate Reservoirs Using a Calcite-Coated Micromodel.* Scientific Reports, 2020. **10**(1).
27. Pak, T., et al., *Droplet fragmentation: 3D imaging of a previously unidentified pore-scale process during multiphase flow in porous media.* Proceedings of the National Academy of Sciences, 2015. **112**(7): p. 1947-1952.
28. Li, J., et al., *Pore-scale investigation of microscopic remaining oil variation characteristics in water-wet sandstone using CT scanning.* Journal of Natural Gas Science and Engineering, 2017. **48**: p. 36-45.
29. Hou, W., et al., *Anti-icing performance of the superhydrophobic surface with micro-cubic array structures fabricated by plasma etching.* Colloids and Surfaces A: Physicochemical and Engineering Aspects, 2020. **586**: p. 124180.
30. Dalawai, S.P., et al., *Recent Advances in durability of superhydrophobic self-cleaning technology: A critical review.* Progress in Organic Coatings, 2020. **138**: p. 105381.
31. Dong, Z., et al., *3D Printing of Superhydrophobic Objects with Bulk Nanostructure.* Advanced Materials, 2021: p. 2106068.
32. Yu, C., et al., *Nature–Inspired self–cleaning surfaces: Mechanisms, modelling, and manufacturing.* Chemical Engineering Research and Design, 2020. **155**: p. 48-65.
33. Yada, S., et al., *Droplet Impact on Surfaces with Asymmetric Microscopic Features.* Langmuir, 2021. **37**(36): p. 10849-10858.

34. Yarin, A.L., *DROP IMPACT DYNAMICS: Splashing, Spreading, Receding, Bouncing....* Annual Review of Fluid Mechanics, 2006. **38**(1): p. 159-192.
35. Sun, C., et al., *Control of water droplet motion by alteration of roughness gradient on silicon wafer by laser surface treatment.* Thin Solid Films, 2008. **516**(12): p. 4059-4063.
36. Lu, Y., et al., *Droplet Directional Movement on the Homogeneously Structured Superhydrophobic Surface with the Gradient Non-Wettability.* Langmuir, 2020. **36**(4): p. 880-888.
37. Grounds, A., R. Still, and K. Takashina, *Enhanced Droplet Control by Transition Boiling.* Scientific Reports, 2012. **2**(1).
38. De Jong, E., et al., *Climbing droplets driven by mechanowetting on transverse waves.* Science Advances, 2019. **5**(6): p. eaaw0914.
39. Zinchenko, A.Z. and R.H. Davis, *A boundary-integral study of a drop squeezing through interparticle constrictions.* Journal of Fluid Mechanics, 2006. **564**: p. 227.
40. Zinchenko, A.Z. and R.H. Davis, *Motion of deformable drops through porous media.* Annual Review of Fluid Mechanics, 2017. **49**: p. 71-90.
41. Wang, G., L. Fei, and K.H. Luo, *Lattice Boltzmann simulation of water droplet impacting a hydrophobic plate with a cylindrical pore.* Physical Review Fluids, 2020. **5**(8): p. 083602.
42. Liu, D., H.-W. Tan, and T. Tran, *Droplet impact on heated powder bed.* Soft matter, 2018. **14**(48): p. 9967-9972.
43. Tan, H., *Three-dimensional simulation of micrometer-sized droplet impact and penetration into the powder bed.* Chemical Engineering Science, 2016. **153**: p. 93-107.
44. Suo, S., M. Liu, and Y. Gan, *An LBM-PNM framework for immiscible flow: With applications to droplet spreading on porous surfaces.* Chemical Engineering Science, 2020. **218**: p. 115577.
45. Markicevic, B., et al., *Infiltration time and imprint shape of a sessile droplet imbibing porous medium.* Journal of Colloid and Interface Science, 2009. **336**(2): p. 698-706.
46. Suo, S. and Y. Gan, *Rupture of Liquid Bridges on Porous Tips: Competing Mechanisms of Spontaneous Imbibition and Stretching.* Langmuir, 2020. **36**(45): p. 13642-13648.
47. Das, S., et al., *Droplet spreading and capillary imbibition in a porous medium: A coupled IB-VOF method based numerical study.* Physics of Fluids, 2018. **30**(1): p. 012112.
48. Shams, M., et al., *A numerical model of two-phase flow at the micro-scale using the volume-of-fluid method.* Journal of Computational Physics, 2018. **357**: p. 159-182.
49. Mirjalili, S., C.B. Ivey, and A. Mani, *Comparison between the diffuse interface and volume of fluid methods for simulating two-phase flows.* International Journal of Multiphase Flow, 2019. **116**: p. 221-238.
50. Brackbill, J.U., D.B. Kothe, and C. Zemach, *A continuum method for modeling surface tension.* Journal of Computational Physics, 1992. **100**(2): p. 335-354.
51. Linder, N., *Numerical Simulation of Complex Wetting*, in *Maschinenbau*. 2015, Technische Universität: Darmstadt.
52. Golpaygan, A. and N. Ashgriz, *Effects of oxidant fluid properties on the mobility of water droplets in the channels of PEM fuel cell.* International Journal of Energy Research, 2005. **29**(12): p. 1027-1040.

53. Golpaygan, A. and N. Ashgriz, *Multiphase flow model to study channel flow dynamics of PEM fuel cells: deformation and detachment of water droplets.* International Journal of Computational Fluid Dynamics, 2008. **22**(1-2): p. 85-95.
54. Ding, H., M.N.H. Gilani, and P.D.M. Spelt, *Sliding, pinch-off and detachment of a droplet on a wall in shear flow.* Journal of Fluid Mechanics, 2010. **644**: p. 217-244.
55. Ghanbarian, B., et al., *Tortuosity in Porous Media: A Critical Review.* Soil Science Society of America Journal, 2013. **77**(5): p. 1461-1477.